\documentclass[12pt]{iopart}
\usepackage{amsfonts}
\usepackage{graphicx}
\eqnobysec
\newcommand{\HA}{\mathcal{H}_{A}}
\newcommand{\HB}{\mathcal{H}_{B}}

\newcommand{\ot}{{\,\otimes\,}}

\newcommand{{\Cn}}{{\mathbb{C}^n}}

\def\<{\langle}
\def\>{\rangle}

\newtheorem{thm}{Theorem}[section]
\newtheorem{Proposition}{Proposition}[section]

\newtheorem{Ex}{Example}[section]
\newtheorem{Remark}{Remark}[section]
\newtheorem{cor}{Corollary}[section]
\newtheorem{Lemma}{Lemma}[section]

\begin{document}


\title[A class of symmetric Bell diagonal entanglement witnesses]{A class of symmetric Bell diagonal entanglement witnesses -- a geometric perspective}

\author{Dariusz  Chru\'sci\'nski}

\address{Institute of Physics, Faculty of Physics, Astronomy and Informatics \\ Nicolaus Copernicus University,\\
Grudzi{a}dzka 5/7, 87--100 Torun, Poland}
\ead{darch@fizyka.umk.pl}



\begin{abstract}
We provide a class of Bell diagonal entanglement witnesses displaying an additional local symmetry -- a maximal commutative subgroup of the unitary group $U(n)$. Remarkably, this class of witnesses is parameterized by a torus being a maximal  commutative subgroup of an orthogonal group $SO(n-1)$. It is shown that a generic element from the class defines an indecomposable entanglement witness.
The paper provides a geometric perspective for some aspects of the entanglement theory and an interesting interplay between group theory and block-positive operators in $\mathbb{C}^n \ot \mathbb{C}^n$.
\end{abstract}


\section{Introduction}

Symmetry plays a prominent role in modern physics. In many cases
it enables one to simplify the analysis of the corresponding
problems and very often it leads to much deeper understanding and
the most elegant mathematical formulation of the corresponding
physical theory. In entanglement theory \cite{QIT,HHHH} the idea
of symmetry was first applied by Werner \cite{WERNER} to
construct an important family of bipartite $n \ot n$ quantum
states which are invariant under the following local unitary
operations
\begin{equation}\label{W}
\rho \ \longrightarrow\  U\ot U \, \rho\, (U \ot U)^\dag\ ,
\end{equation}
for any $U\in U(n)$, where $U(n)$ denotes the group of unitary $d
\times d$ matrices. Another family of symmetric states (so called
isotropic states) is governed by the following
invariance rule
\begin{equation}\label{I}
\rho\  \longrightarrow\  U\ot {U}^* \, \rho \, (U \ot {U}^*)^\dag\ ,
\end{equation}
where ${U}^*$ is the complex conjugate of $U$ in some fixed orthonormal basis $\{e_0,\ldots,e_{n-1}\}$ in $\mathbb{C}^n$ (see \cite{Werner2,Keyl}).
If we allow the full unitary group $U(n)$, then the only bipartite operators invariant under $U(n) \ot U(n)$ are the identity operator $\mathbb{I}_n \ot \mathbb{I}_n$ and the flip (or swap) operator $\mathbb{F}$ defined by $\mathbb{F}\, x \ot y = y \ot x$. Similarly, the only bipartite operators invariant under $U(n) \ot U(n)^*$ are the identity operator invariant $\mathbb{I}_n \ot \mathbb{I}_n$ and the rank-1 projector onto the maximally entangled state $P^+_n = |\psi^+_n\>\<\psi^+_n|$, where  $|\psi^+_n\> = \frac{1}{\sqrt{n}} \sum_k e_k \ot e_k$. One finds the following formulae for the Werner state
\begin{equation}\label{}
    \rho_{f} = \frac{1}{n(n-f)}\, (\mathbb{I}_n \ot \mathbb{I}_n - f \mathbb{F}) \ ,
\end{equation}
and isotropic state
\begin{equation}\label{}
    \rho_{p} = \frac{1-p}{n^2} \mathbb{I}_n \ot \mathbb{I}_n + p P^+_n \ ,
\end{equation}
respectively. Remarkably, the properties of these two families of bipartite symmetric states are fully controlled by the operation of partial transposition: both $\rho_{f}$ and $\rho_{p}$ are separable iff they are PPT, i.e. i.e. $f \leq 1/n$ and $p \leq 1/(n+1)$ for Werner and isotropic state, respectively (a bipartite state $\rho$ is PPT if its partial transposition $\rho^\Gamma$ defines a positive operator).
This example shows how symmetry simplifies separability problem
in the entanglement theory. A general separability problem is much harder and the classification
of states of a composite quantum system is very subtle \cite{HHHH,Guhne}. Let us recall that
the most general approach to characterize quantum entanglement uses a notion of an
entanglement witness. A Hermitian operator $W$ in acting in $\HA \ot  \HB$ is block-positive if $\< x \ot y|W|x \ot y \> \geq  0$ for all $x \in \HA$ and $y \in \HB$. Clearly, a positive operator is necessarily block-positive but the converse needs not be true. An entanglement witness (a notion introduced by Terhal \cite{Terhal}) is a block-positive operator which is not positive, i.e. it
possesses at least one negative eigenvalue  (see a recent review \cite{TOPICAL} for detailed presentation). Remarkably, it turns out that any entangled state can be detected by some entanglement witness
and hence the knowledge of witnesses enables us to perform full classification of states
of composite quantum systems: a state $\rho$ living in $\HA \ot  \HB$ is entangled iff there is an
entanglement witness $W$ such that $\tr(\rho W) < 0$ \cite{HHHH}. An entanglement witness $W$ is optimal
\cite{opt} if there is no other witness which detects more entangled states than $W$. In the class
of $U \ot  U$-invariant EWs an optimal witness is provided by a flip operator
\begin{equation}\label{I}
    W = \mathbb{F}\ .
\end{equation}
Similarly, an optimal $U \ot U^*$-invariant EW is provided by
\begin{equation}\label{II}
    W' = \mathbb{I}_n \ot \mathbb{I}_n - nP^+_n \ .
\end{equation}
One easily finds that a Werner state $\rho_f$ is entangled iff $\tr(\mathbb{F} \rho_f ) < 0$ and similarly
an isotropic state $\rho_p$ is entangled iff $\tr(W' \rho_p)  < 0$. Both witnesses \eref{I} and \eref{II} are
decomposable, i.e. $W = A+B^\Gamma$, where $A,B \geq 0$ and $B^\Gamma$ denotes a partial transposition
of $B$. Decomposable EWs can not detect PPT entangled states. It should be stressed
there is no universal method to construct an indecomposable EW which can be used to
detect PPT entangled states.

It is, therefore,  clear that define a bigger class of symmetric  states and entanglement witnesses one has to restrict the local symmetry from the full unitary group $U(n)$ to one of its subgroups. In this paper we consider $G \ot G^*$-invariant bipartite operators in $\Cn \ot \Cn$, where $G$ defines a subgroup of $U(n)$. Within a class of such $G \ot G^*$-invariant operators we provide a detailed analysis of entanglement witnesses. Remarkably, a generic EW from this class in indecomposable and hence it may serve as a detector of bound entanglement. The paper provides a geometric perspective for some aspects of the entanglement theory and an interesting interplay between group theory and block-positive operators in $\mathbb{C}^n \ot \mathbb{C}^n$.

\section{A class of symmetric operators}

Let us consider the following subgroup
\begin{equation}\label{}
    G_1 = \{\ U \in U(n)\ | \ U=\sum_{k=0}^{n-1} e^{i\phi_k}\, E_{kk}\ \} \subset U(n)\ ,
\end{equation}
where $E_{kl} := |e_k\>\<e_l|$ and $\phi_k \in [0,2\pi)$. Note, that $G_1$ is a maximal commutative
subgroup of $U(n)$ ($n$-dimensional torus parameterized by angles $\phi_k$). Now, the  $(G_1 \ot G_1^*)$-invariant operator has the following form \cite{torus}
\begin{equation}\label{X1}
    X = \sum_{k,l=0}^{n-1} a_{kl} E_{kk} \ot E_{ll}  + \sum_{k\neq l=0}^{n-1} b_{kl} E_{kl} \ot E_{kl}  \ .
\end{equation}
Note, that $X$ is Hermitian iff $a_{kl} \in \mathbb{R}$ and $b_{kl} = b_{lk}^*$. Consider now a discrete subgroup
\begin{equation}\label{}
    G_2 = \{ \ \lambda^m U_{kl} \ | \ k,l,m=0,1,\ldots,n-1 \} \subset U(n)\ ,
\end{equation}
where  $   \lambda= e^{2\pi i/d}$  and $U_{kl}$ denotes a family of  unitary Weyl operators defined as follows \cite{Hiesmayr1,Bertlman1,Weyl}
\begin{equation}\label{U_mn}
    U_{mk} e_l = \lambda^{ml} e_{l+k}\ , \ \ \ \ {\rm mod}\ n \ .
\end{equation}
The matrices $U_{kl}$ satisfy
\begin{equation}\label{}
  U_{kl} U_{rs} = \lambda^{ks} U_{k+r,l+s} \ , \ \ \ U_{kl}^* = U_{-k,l}\ , \ \ \ U_{kl}^\dagger = \lambda^{kl} U_{-k,-l}\ ,
\end{equation}
and the following orthogonality relations
\begin{equation}\label{}
    {\rm tr}(U_{kl} U_{rs}^\dagger) = n\, \delta_{kr} \delta_{ls} \ .
\end{equation}
One has therefore
\begin{equation}\label{}
    G_2 \ot G_2^* = \{ \  U_{kl} \ot U_{-k,l} \ | \ k,l=0,1,\ldots,n-1 \} \ .
\end{equation}
Note, that $G_2 \ot G_2^*$ defines a discrete commutative subgroup of $U(n) \ot U^*(n)$. Interestingly, its commutant, that is, an algebra of $G_2 \ot G_2^*$-invariant operators is spanned by $U_{kl} \ot U_{-k,l}$ and hence any $G_2 \ot G_2^*$-invariant operator has the following form
\begin{equation}\label{X2}
    X = \sum_{k,l=0}^{n-1} c_{kl}\, U_{kl} \ot U_{-k,l} \ .
\end{equation}
Note, that \eref{X2} defines a Hermitian operator iff
\begin{equation}\label{cc}
    c_{kl} = c_{n-k,n-l}^*\ .
\end{equation}
Denote by $|\psi_{kl}\>$  generalized Bell states in $\mathbb{C}^n \ot \mathbb{C}^n$
\begin{equation}\label{}
  |\psi_{kl}\> = \mathbb{I}_n \ot U_{kl} |\psi^+_n\> \ ,
\end{equation}
and let $P_{kl} = |\psi_{kl}\>\<\psi_{kl}|$ be the corresponding rank-1 projectors. One easily shows that $P_{kl}$ span the entire commutant of $G_2 \ot G_2^*$ and hence any $G_2 \ot G_2^*$-invariant operator is Bell diagonal, that is, it can be represented as follows
\begin{equation}\label{}
    X = \sum_{k,l=0}^{n-1} x_{kl} P_{kl}\ .
\end{equation}
One easily finds

\begin{Lemma} \label{LI}
A Hermitian $G_1\ot G_1^*$-invariant operator \eref{X1} is $G_2 \ot G_2^*$-invariant if the matrix $a_{kl}$ is circulant, that is,  $a_{kl} = \alpha_{k-l} \in \mathbb{R}$, and $b_{kl} = c \in \mathbb{R}$.
\end{Lemma}
Similarly,
\begin{Lemma} \label{LII}
A Hermitian $G_2\ot G_2^*$-invariant operator \eref{X2} is $G_1 \ot G_1^*$-invariant if the matrix $c_{kl}$ has the following structure
\begin{equation}\label{}
    c_{kl} = \left( \begin{array}{cccc} c_0 & c & \ldots & c \\ c_1 & c & \ldots & c \\ \vdots & \vdots & \ddots &\vdots \\
    c_{n-1} & c & \ldots & c \end{array} \right) \ ,
\end{equation}
where `$c$'  is an arbitrary real parameter and the vector $c_k := c_{k0}$ is defined as follows
\begin{equation}\label{c-a}
    c_{k} = \sum_{l=0}^{n-1} \omega^{-kl} \alpha_l\ ,
\end{equation}
that is, it is a discrete Fourier transform of a real vector $\alpha_l$.
\end{Lemma}
It is, therefore, clear that two representations \eref{X1} and \eref{X2} are complementary to each other.
Now, combining \eref{X1} and \eref{X2} we obtain the following
formula for a spectral resolution of any $G_1 \ot G_1^*$-invariant Bell diagonal operator
\begin{equation}\label{X3}
    X =  (\alpha_0+1) \Pi_0 + \sum_{k=1}^{n-1} \alpha_k \Pi_k + \beta\,  n P^+_n  \ ,
\end{equation}
where
\begin{equation}\label{}
    \Pi_k = P_{0k} + P_{1k} + \ldots + P_{n-1,k} \ ,  \ \ \ \ k=0,\ldots,n-1\ .
\end{equation}
Now, if \eref{X3} represents an EW then necessarily $\alpha_k \geq 0$ for $k=0,\ldots,n-1$ and $\beta < 0$.
From now on we fix  $\beta = -1$. Clearly, these conditions are necessary but not sufficient.
We pose the following question: what are the additional properties of $\{\alpha_0,\ldots,\alpha_{n-1}\}$  which
guarantee that the formula \eref{X3} provides a legitimate entanglement witness. Note,
that if $\alpha_0=0$ and $\alpha_1=\ldots=\alpha_{n-1}=1$, then \eref{X3} reconstructs \eref{II}. The class of
witnesses
\begin{equation}\label{EW}
    W[\alpha_0,\ldots,\alpha_{n-1}] :=  (\alpha_0+1) \Pi_0 + \sum_{k=1}^{n-1} \alpha_k \Pi_k - n P^+_n  \ ,
\end{equation}
seems to be very special, however, it turns out that many EWs considered in
the literature belong to this class.

\section{Entanglement witnesses vs. positive maps}

Due to the Choi-Jamio{\l}kowski isomorphism any entanglement witness $W$ in $\Cn \ot \Cn$
corresponds to a positive map $\Lambda : M_n(\mathbb{C}) \rightarrow  M_n(\mathbb{C})$ {\em via} the following relation
\begin{equation}\label{CJ}
    W = \sum_{i,j=0}^{n-1} E_{ij} \ot \Lambda(E_{ij})\ .
\end{equation}
The map corresponding to \eref{EW} has the following form
\begin{eqnarray}  \label{Lambda}
    \Lambda(E_{ii}) &=& \sum_{j=0}^{n-1} a_{ij} E_{jj}\ , \nonumber \\
    \Lambda(E_{ij}) &=& - E_{ij} \ , \ \ \ i \neq j\ ,
\end{eqnarray}
where $a_{ij} := \alpha_{i-j} \geq 0$.

\begin{Proposition} A linear map $\Lambda$ is positive if and only if the following cyclic inequalities
\begin{equation}\label{CYC}
  \sum_{i=0}^{n-1} \frac{t_i^2}{(\alpha_0 + 1)t_i^2 + \sum_{k=1}^{n-1} \alpha_k t_{i+k}^2 } \leq 1\ .
\end{equation}
are satisfied for all $\,t_0, t_1 , \ldots, t_n \geq 0$. $\Lambda$ is completely positive if and only if $\alpha_0 \geq n-1$.
\end{Proposition}
In particular taking $t_0 = \ldots = t_{n-1}$ one finds
\begin{equation}\label{n-1}
  \alpha_0 + \alpha_1 + \ldots + \alpha_{n-1} \geq n-1\ .
\end{equation}
Hence, if $ W[\alpha_0,\ldots,\alpha_{n-1}]$ is an entanglement witness, then necessarily
$\{\alpha_0,\ldots,\alpha_{n-1}\}$ satisfy \eref{n-1} and additionally
\begin{equation}\label{0-n}
    0 \leq \alpha_0 < n-1\ .
\end{equation}
Interestingly, one has
\begin{Proposition} For $\, n = 2$ conditions \eref{n-1} and \eref{0-n} are necessary and sufficient.
\end{Proposition}
However, for $n \geq 3$ these conditions  are no longer sufficient. For $n = 3$
introducing $a = \alpha_0$, $b = \alpha_1$ and $c = \alpha_2$ one has the following well known result

\begin{thm}[\cite{Cho-abc}]
An operator $W[a,b,c]$ is an entanglement witness if and only if apart
from \eref{n-1} and \eref{0-n} the following extra condition has to be satisfied: if $a \leq 1$, then
\begin{equation}\label{k1}
    bc \geq (1-a)^2 \ .
\end{equation}
Moreover, being  an entanglement witness it is indecomposable if and only if
\begin{equation}\label{k2}
    4bc < (2-a)^2\ .
\end{equation}
\end{thm}
From now on we consider entanglement witnesses $W[\alpha_0,\ldots,\alpha_{n-1}]$ which belong to the boundary of a set of entanglement witnesses. Clearly, any optimal witness belongs to this boundary. Note, that the corresponding parameters $\{\alpha_0,\ldots,\alpha_{n-1}\}$ instead of \eref{n-1} satisfy the following equality
\begin{equation}\label{==}
  \alpha_0 + \alpha_1 + \ldots + \alpha_{n-1} = n-1\ .
\end{equation}
Now, for $n = 3$  we look for a set of parameters $a,b,c \geq 0$ belonging to a simplex $a + b + c = 2$ and satisfying for $a \leq 1$
\begin{equation}\label{}
    bc = (1-a)^2 \ ,
\end{equation}
which corresponds to the boundary of a set defined by an inequality  \eref{k1}. Actually, the above condition defines an ellipse $bc = (b + c -1)^2$ on the $bc$-plane (cf. \cite{FilipI}). It is easy to show that the above conditions, i.e.
\begin{equation}\label{}
    a+b+c=2\ , \ \ \ bc = (1-a)^2 \ ,
\end{equation}
are equivalent to much more symmetric ones
\begin{equation}\label{EW-3}
    a+b+c=2\ , \ \ \ a^2 + b^2 + c^2 = 2 \ .
\end{equation}
Now, the intersection of the 2D sphere $a^2+b^2+c^2 = 2$ and the plane $a+b+c = 2$ defines
a circle and its projection on the $bc$-plane gives rise to an ellipse $bc = (b + c - 1)^2$ (cf.
Fig. 1). Note, that equivalently one may describe the above circle as an intersection of
the following sphere centered at $(1,1,1)$
\begin{equation}\label{S3a}
(a - 1)^2 + (b - 1)^2 + (c - 1)^2 = 1\ ,
\end{equation}
or the one centered at $(\frac 23,\frac 23,\frac 23)$ (the middle of the simplex)
\begin{equation}\label{S3b}
    \left( a - \frac 23 \right)^2 +\left( b - \frac 23 \right)^2+\left( c - \frac 23 \right)^2 = \frac 23 \ ,
\end{equation}
with a plane $a+b+c=2$.

\begin{figure}[!h] \label{FIG}
 \includegraphics[width=5cm]{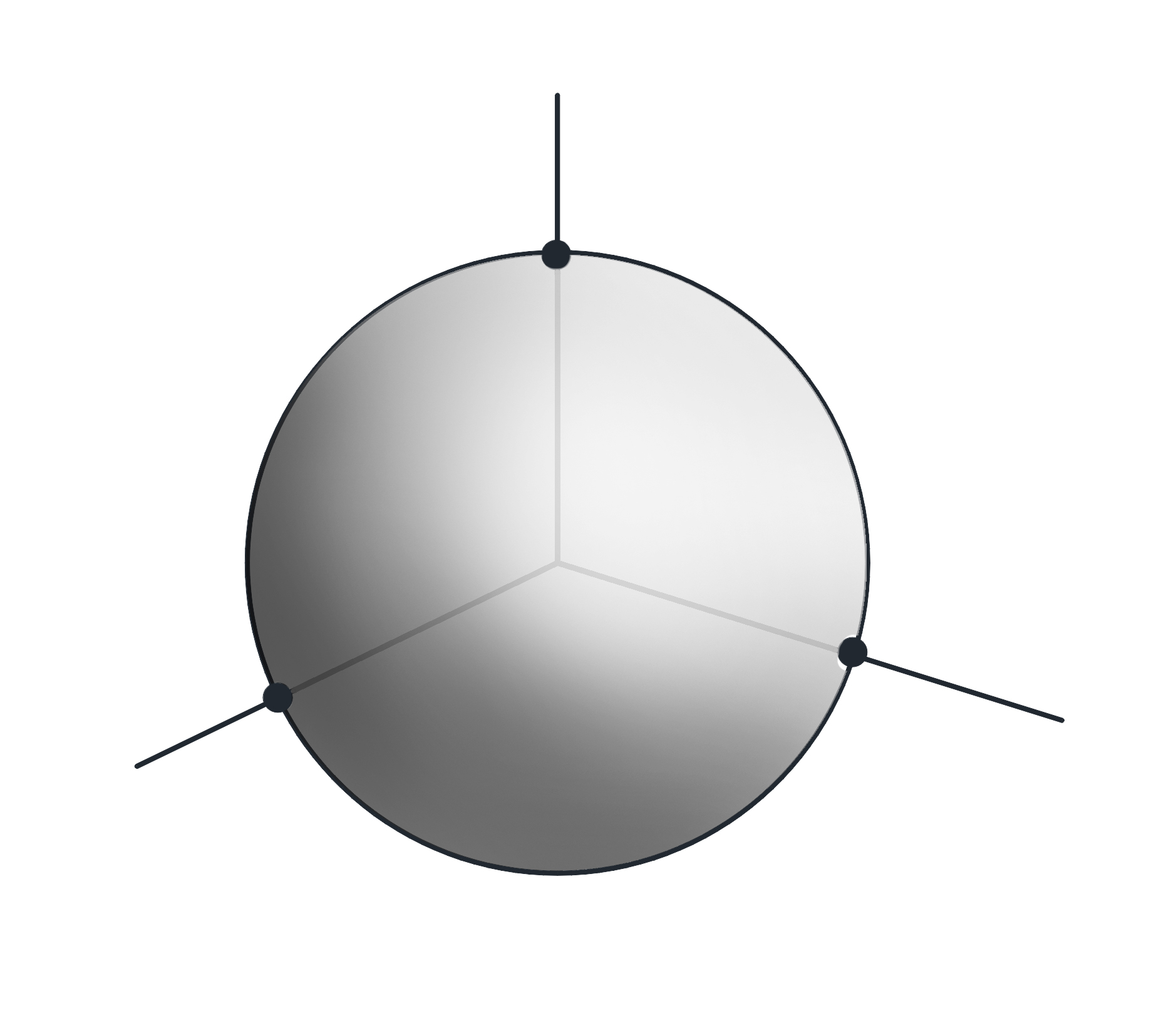}
 \includegraphics[width=5.5cm]{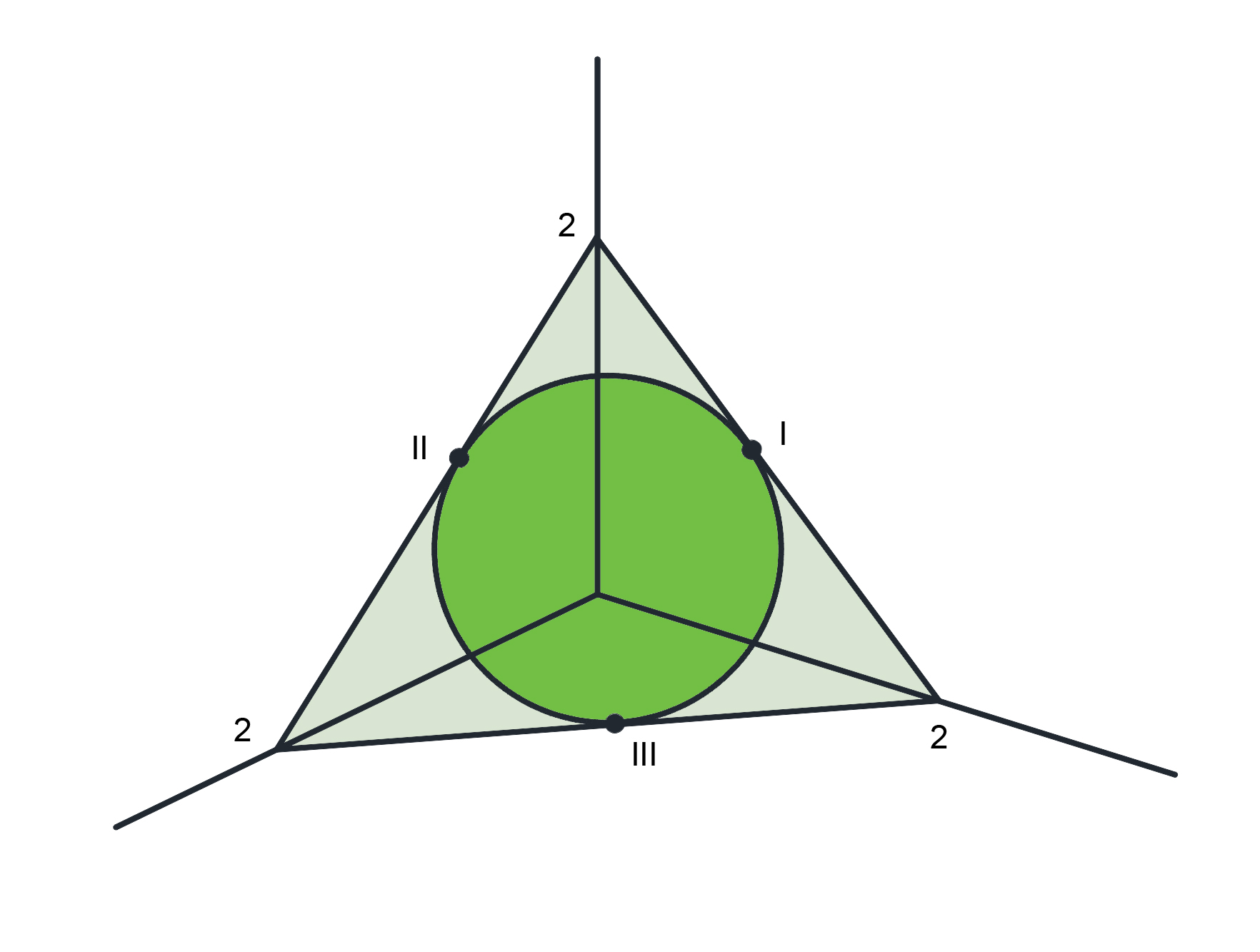}
 \includegraphics[width=5cm]{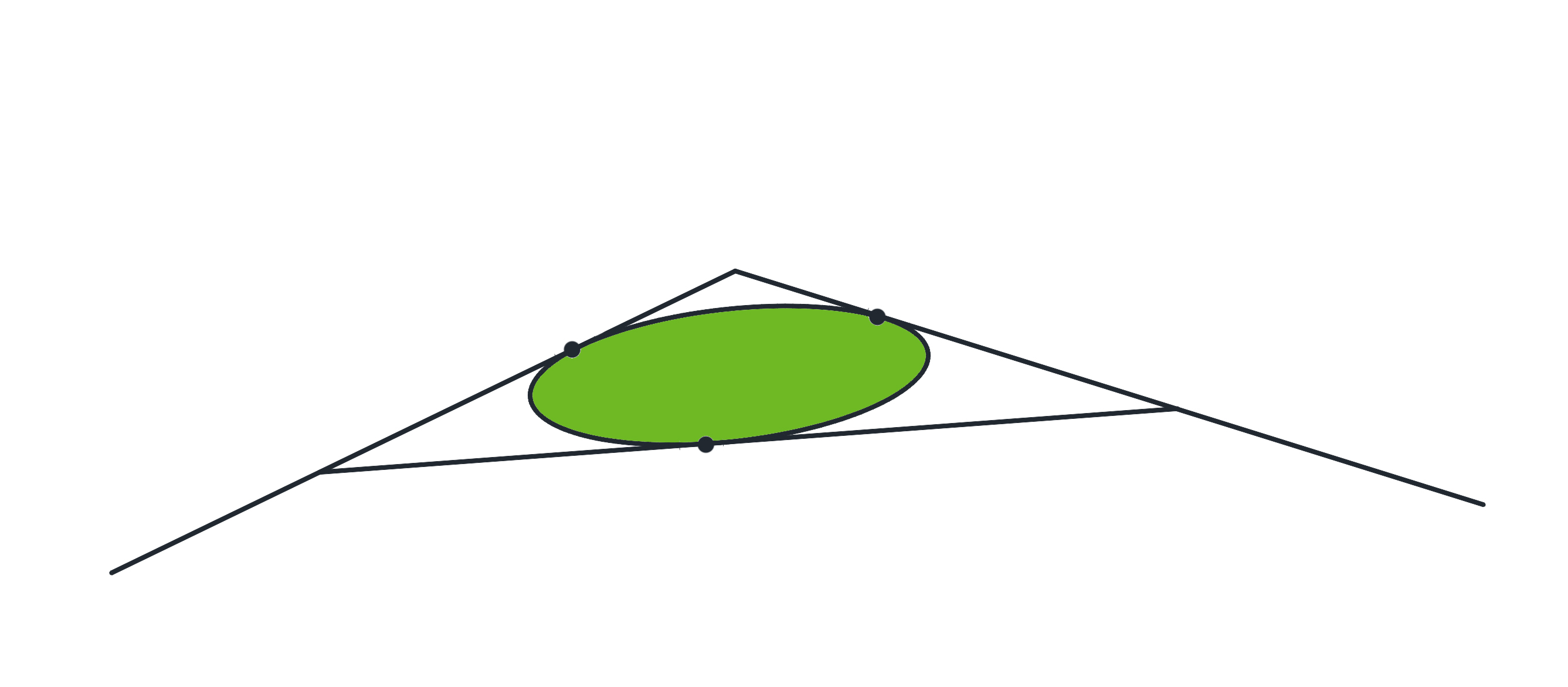}
 \caption{On the left: a 2D sphere $a^2 + b^2 + c^2 = 2$. On the middle: the intersection
of a sphere and a simplex $a + b + c = 2$. On the right: an ellipse on a $bc$-plane being a projection
of a circle. Characteristic points: I and II correspond to Choi maps and III to the
reduction map.}
 \end{figure}
It should be stressed that for $n > 3$ we do not know the complete set of conditions implied by the cyclic
inequalities \eref{CYC} (see \cite{FilipII} for partial results for $n = 4$).

\section{Witnesses parameterized by an orthogonal group}

In this section we analyze a class of witnesses $ W[\alpha_0,\ldots,\alpha_{n-1}]$ generated by a certain
family of positive maps proposed in \cite{Kossak-kule}: let us define a set of Hermitian traceless matrices
\begin{equation}\label{}
    F_\ell  = \frac{1}{\sqrt{\ell(\ell +1)}} \Big( \sum_{k=0}^{\ell-1} E_{kk} - \ell E_{\ell\ell} \Big) \ ,\ \ \ \ \ell = 1,\ldots,n-1\ .
\end{equation}
One defines a real $n \times n$ matrix
\begin{equation}\label{aR}
    a_{ij} = \frac{n-1}{n} + \sum_{\alpha,\beta=1}^{n-1} \< e_i|F_\alpha|e_i\> R_{\alpha\beta} \<e_j|F_\beta|e_j\> \ ,
\end{equation}
where $R_{\alpha\beta}$ is an orthogonal $(n-1)\times (n-1)$ orthogonal matrix. Due to the fact that $F_\alpha$ is traceless
for $\alpha=1,\ldots,n-1$, one finds
\begin{equation}\label{doubly}
    \sum_{i=1}^{n-1} a_{ij} = \sum_{j=1}^{n-1} a_{ij} = n-1\ ,
\end{equation}
Moreover, it turns out \cite{Kossak-kule} that matrix elements $a_{ij} \geq  0$ and hence
\begin{equation}\label{}
    \widetilde{a}_{ij} := \frac{1}{n-1}\, a_{ij}\ ,
\end{equation}
defines a doubly stochastic matrix. Consider now a linear map $\Lambda$ defined by \eref{Lambda} with
$a_{ij}$  defined by \eref{aR}.

\begin{Proposition}[\cite{Kossak-kule}] For any orthogonal matrix $R_{\alpha\beta}$ a linear map $\Lambda$ is positive.
\end{Proposition}
Suppose we are given a $n \times n$ matrix $a_{ij}$ such that $a_{ij} \geq  0$ and \eref{doubly} is satisfied.

\begin{Proposition}[\cite{OSID}] A matrix $a_{ij}$ can be represented by \eref{aR} if and only if
\begin{equation}\label{osid-a}
    \sum_{k=0}^{n-1} a_{ik} a_{jk} = \delta_{ij} + n-2\ ,
\end{equation}
for $\, i,j=0,\ldots,n-1$.
\end{Proposition}
Hence, if the matrix $a_{ij}$ is circulant, i.e. $a_{ij} = \alpha_{i-j}$, then  \eref{osid-a} implies the following set of conditions for a set of parameters $\{\alpha_0,\ldots,\alpha_{n-1}\}$:
\begin{equation}\label{osid-b}
    \sum_{k=0}^{n-1} \alpha_{i-k} \alpha_{j-k} = \delta_{ij} + n-2\ .
\end{equation}

\begin{Ex} For $\, n= 3$ using again the following notation $a = \alpha_0$, $b = \alpha_1$ and $c = \alpha_2$ the formula \eref{osid-b} implies
\begin{equation}\label{3-1}
    a^2 + b^2 + c^2 = 2 \ , \ \ \ {\rm for}\  i = j  \ ,
\end{equation}
and
\begin{equation}\label{3-2}
ac + ba + cb = 1 \ ,  \ \ \ {\rm for}\ i \neq j \ .
\end{equation}
Note, however, that \eref{3-1} and \eref{3-2} are not independent. Indeed, taking into account $a+b+c=2$ one has
\begin{equation*}\label{}
    4 = (a+b+c)^2 =  a^2 + b^2 + c^2 + 2(ac + ba + cb) \ ,
\end{equation*}
and hence \eref{3-1} implies \eref{3-2}. One concludes, therefore, that this class is fully characterized by
\begin{equation}\label{}
    a+b+c=2\ , \ \ \ a^2 + b^2 + c^2 = 2 \ ,
\end{equation}
which reproduce \eref{EW-3}.
\end{Ex}

\begin{Ex} \label{E-n=4} For $\, n= 4$ using  $a = \alpha_0$, $b = \alpha_1$, $c = \alpha_2$ and $d=\alpha_3$
one has
\begin{equation}\label{4-0}
    a+b+c+d=3\ ,
\end{equation}
and the formula \eref{osid-b} implies
\begin{equation}\label{3-1}
    a^2 + b^2 + c^2 +d^2 = 3 \ , \ \ \ ac + bd = 1  , \ \ \ (a + c)(b + d) = 2 \ .
\end{equation}
Actually, assuming \eref{4-0} only two of the above three conditions are independent.
Introducing $x = a + c$ and $y = b + d$ one obtains the following equations for a pair
$(x,y)$:
\begin{equation*}\label{}
    xy = 2 \ , \ \ \  x^2 + y^2 = 5 \ ,
\end{equation*}
with two solutions $(x = 1, y = 2)$ and $(x = 2, y = 1)$. Finally, we have two classes of
admissible parameters $\{a,b,c,d\}$ constrained by
\begin{equation}\label{4I}
    a + b + c + d = 3 \ , \ \  a^2 + b^2 + c^2 + d^2 = 3 \ , \ \  b + d = 1  \ ,
\end{equation}
and
\begin{equation}\label{4II}
    a + b + c + d = 3 \ , \ \  a^2 + b^2 + c^2 + d^2 = 3 \ , \ \  b + d = 2\ .
\end{equation}
Note, that  the intersection of a 3D sphere $a^2 + b^2 + c^2 + d^2 = 3$ with a simplex $a + b + c + d = 3$ may be equivalently rewritten as
the intersection with the following sphere centered at $(1, 1, 1, 1)$
\begin{equation}\label{S4a}
(a - 1)^2 + (b - 1)^2 + (c - 1)^2 + (d - 1)^2 = 1 \ ,
\end{equation}
or the one centered at the middle of the simplex $( \frac 34,\frac 34,\frac 34,\frac 34)$
\begin{equation}\label{S4b}
    \left( a - \frac 34 \right)^2 +\left( b - \frac 34 \right)^2+\left( c - \frac 34 \right)^2 + \left( d - \frac 34 \right)^2 = \frac 34 \ ,
\end{equation}
which provide analogs of \eref{S3a} and \eref{S3b}, respectively.
\end{Ex}
Clearly, for higher dimensions the number of conditions implied by \eref{osid-b} grows: one
always has
\begin{equation}\label{SS}
    \sum_{k=0}^{n-1} \alpha_k^2 = n-1\ ,
\end{equation}
corresponding to $i = j$ plus some extra conditions following from \eref{osid-b} for $i \neq j$. The
above conditions define $(n- 2)$-dim. sphere as an intersection of $(n - 1)$-dim. sphere \eref{SS} with the simplex $\sum_{k=0}^{n-1} \alpha_k = n-1$. The same intersection is provided by
\begin{equation}\label{SS}
    \sum_{k=0}^{n-1} [\alpha_k-1]^2 = 1\ ,
\end{equation}
and
\begin{equation}\label{SS}
    \sum_{k=0}^{n-1} \left( \alpha_k^2 - \frac{n-1}{n}\right)^2 = \frac{n-1}{n}\ ,
\end{equation}
in analogy with \eref{S4a} and \eref{S4b}, respectively.

\section{Witnesses constructed from Weyl operators}

Now, we provide characterization of entanglement witnesses from the previous section using a complementary representation \eref{X2}.
Authors of \cite{Weyl} provided the following

\begin{Proposition} Let $W$ be a Hermitian Bell diagonal operator defined by
\begin{equation}\label{Wa}
    W = a\,\sum_{k,l=0}^{n-1} c_{kl}\, U_{kl} \ot U_{-k,l} \ ,
\end{equation}
with $a>0$, $c_{00} = n-1$.  If  $|c_{kl}| \leq 1$ (apart from $c_{00}$), then $W$ is block positive.
\end{Proposition}
Using Lemma \ref{LII} one easily finds that formula \eref{Wa} reproduces $W = W[\alpha_0,\ldots,\alpha_{n-1}]$ iff
\begin{equation}\label{}
    a = \frac 1n\ , \ \ \ c_{kl} = 1 \ , \ \ \ l=1,\ldots,n-1\ ,
\end{equation}
and
\begin{equation}\label{DFT}
    c_{k0} = \sum_{l=0}^{n-1} \omega^{-kl} \alpha_l\ .
\end{equation}
Note that formula \eref{DFT} implies $c_{00} = \alpha_0 + \ldots + \alpha_{n-1} = n-1$. Interestingly, a set of conditions  \eref{osid-b} for parameters $\alpha_k$ is equivalent to  a set of remarkably simple conditions for parameters $c_{k0}$.

\begin{Proposition} A set $\{\alpha_0,\ldots,\alpha_{n-1}\}$  such that $\alpha_k \geq 0$ and $\alpha_0 + \ldots \alpha_{n-1} = n-1$
satisfies \eref{osid-b} if and only if a set of $c_{k0}$ defined by \eref{DFT} satisfies
\begin{equation}\label{}
 c_{00} = n-1\ , \ \ \ \   c_{k0} = c_{n-k,0}^*\ , \ \ \ \ |c_{k0}| = 1 \ ,
\end{equation}
for $k=1,\ldots,n-1$.
\end{Proposition}
Proof:  one has
\begin{equation}\label{}
     c_{n-k,0} = \sum_{l=0}^{n-1} \omega^{-(n-k)l} \alpha_l = \sum_{l=0}^{n-1} \omega^{-nl} \omega^{kl} \alpha_l = c_{k0}^*\ ,
\end{equation}
due to $\omega^{nl} =1$. Now, the inverse to \eref{DFT} reads
\begin{equation}\label{DFT-1}
    \alpha_{k} = \frac 1n \sum_{l=0}^{n-1} \omega^{kl} c_{l0}\ .
\end{equation}
Suppose now that $|c_{k0}|=1$. Using the fact that $\alpha_k = \alpha_k^*$
one has
\begin{eqnarray}
 \sum_{k=0}^{n-1} \alpha_k \alpha_k^*  &=& \frac{1}{n^2} \sum_{k=0}^{n-1} \sum_{i=0}^{n-1} \omega^{ik} c_{i0}   \sum_{j=0}^{n-1} \omega^{-jk} c_{j0}  = \frac{1}{n^2} \sum_{i,j=0}^{n-1}\Big(  \sum_{k=0}^{n-1} \omega^{(i-j)k} \Big) c_{i0}c_{j0}^* \nonumber   \\
   &=& \frac 1n  \sum_{i=0}^{n-1} |c_{i0}|^2 = \frac 1n [ (n-1)^2 + (n-1)] = n-1\ ,
\end{eqnarray}
where we have used
\begin{equation}\label{}
    \sum_{k=0}^{n-1} \omega^{(i-j)k} = n \delta_{ij}\ .
\end{equation}
Similarly, for $i \neq j$
\begin{eqnarray}
 \sum_{k=0}^{n-1} \alpha_{i-k} \alpha_{j-k}^*    &=& \frac{1}{n^2} \sum_{k=0}^{n-1} \sum_{r=0}^{n-1} \omega^{r(i-k)} c_{r0}   \sum_{s=0}^{n-1} \omega^{-s(j-k)} c_{s0}^* \nonumber  \\
   &=& \frac{1}{n^2} \Big(  \sum_{k=0}^{n-1} \omega^{(s-r)k} \Big) \sum_{r,s=0}^{n-1} \omega^{ri} c_{r0}  \omega^{-sj} c_{s0}^*
     = \frac{1}{n} \sum_{r=0}^{n-1} \omega^{r(i-j)} |c_{r0}|^2  \nonumber  \\
     &=& \frac 1n ( [n-1]^2 - 1 ) = n-2\ ,
\end{eqnarray}
which proves \eref{osid-b}. Conversely, if \eref{osid-b} is satisfies, then in a similar way one shows that $|c_{k0}|^2=1$. \hfill $\Box$

Hence the entire class of witnesses is parameterized by phases  of $c_{k0} = e^{i\phi_k}$. Due to $c_{k0}= c_{n-k,0}^*$ one has two cases:

\begin{enumerate}

\item if $n=2m+1$, then we have $m$ independent phases $c_{10} = e^{i\phi_1}, \ldots , c_{m0} = e^{i\phi_m}$.

\item if $n=2m+2$, then we have $m$ independent phases $c_{10} = e^{i\phi_1}, \ldots , c_{m0} = e^{i\phi_m}$ and one real parameter $c_{m+1,0} = \pm 1$.

\end{enumerate}
It shows that for an odd $n$ ($n=2m+1$) the space of witnesses is parameterized by $m$-dim. torus $\mathbb{T}_m$ and if $n$ is even $(n=2m+2)$ we have two classes of witnesses: each one corresponding to $\mathbb{T}_m$.

\begin{Remark} A similar observation holds for PPT Bell diagonal states, i.e. the structure of PPT Bell diagonal states in $\Cn \ot \Cn$  depends upon the parity of `$n$' (cf. \cite{Hiesmayr1}).
\end{Remark}

\begin{Ex} For $n=3$ putting $c_{10} = e^{i\phi}= c_{20}^*$ one finds
\begin{eqnarray}\label{}
    a &=& \frac 13 ( 2 +  c_{10} + c_{10}^*) = \frac 23 ( 1 + \cos\phi) \ , \nonumber \\
    b &=& \frac 13 ( 2 + \omega c_{10} + \omega^*c_{10}^*) = \frac 13 ( 2 - \cos\phi - \sqrt{3} \sin\phi) \ , \\
    c &=& \frac 13 ( 2 + \omega^* c_{10} + \omega c_{10}^*) = \frac 13 ( 2 - \cos\phi + \sqrt{3} \sin\phi) \nonumber \ ,
\end{eqnarray}
due to $\omega =e^{2\pi i/3} = \frac 12 (-1 + i \sqrt{3})$.
\end{Ex}

\begin{Ex} For $n=4$ if $c_{10} = e^{i\phi}= c_{30}^*$ and $c_{20} = 1$ one finds
\begin{equation}\label{}
\fl    a = \frac 12(2 + \cos\phi) \ , \ \  b = \frac 12(1 - \sin\phi) \ , \ \ c = \frac 12(2 - \cos\phi) \ , \ \ d = \frac 12 (1 + \sin\phi) \ ,
\end{equation}
and similarly if $c_{10} = e^{i\psi}= c_{30}^*$ and $c_{20}=-1$ one has
\begin{equation}\label{}
 \fl   a = \frac 12(1 + \cos\psi) \ , \ \  b = \frac 12(2 - \sin\psi) \ , \ \ c = \frac 12(1 - \cos\psi) \ , \ \ d = \frac 12 (2 + \sin\psi) \ .
\end{equation}
Note, that for $c_{20}=1$ one has $b+d=1$, whereas for $c_{20}=-1$ one has $b+d=2$. This way we reproduced two classes from Example \ref{E-n=4}.
\end{Ex}

It should be clear that the structure of tori $\mathbb{T}_m$ is related with the properties of orthogonal group considered in the previous section. Note, that the structure of the orthogonal group differs in certain aspects between even and odd dimensions. For example, the reflection corresponding to `$-\mathbb{I}$' is orientation-preserving in even dimensions, but orientation-reversing in odd dimensions.

\begin{enumerate}

\item If $n=2m+1$, then $O(n-1) = O(2m)$ and a single torus $\mathbb{T}_m$ corresponds to a maximal commutative subgroup of $SO(2m)$.

\item If $n=2m+2$, then $O(n-1) = O(2m+1)$ and we have two tori $\mathbb{T}_m$ and $\mathbb{T}_m'$. Torus $\mathbb{T}_m$ corresponds to a maximal commutative subgroup of $SO(2m+1)$ whereas $\mathbb{T}_m'$ is defined by composing $\mathbb{T}_m$  with reflection, that is, $g \in \mathbb{T}_m'$ iff $-g \in \mathbb{T}_m$.

\end{enumerate}

\begin{Remark} It should be stressed that a set $\{\alpha_0,\ldots,\alpha_{n-1}\}$ satisfying \eref{osid-b} provides only
a proper subset of admissible parameters. Note that
\begin{equation}\label{}
    \alpha_0 \leq 2\,\frac{n-1}{n} < 2\ ,
\end{equation}
and hence one can not reproduce well known entanglement witnesses corresponding to
\begin{equation}\label{}
    \alpha_0 = n-k\ , \ \alpha_1 = \ldots = \alpha_{k-1}=1\ , \ \alpha_k = \ldots = \alpha_{n-1} = 0\ ,
\end{equation}
for $k=2,\ldots,n-2$.
\end{Remark}

\section{Decomposability and optimality}

Finally, we address the problem of decomposability of $W[\alpha_0,\ldots,\alpha_{n-1}]$.

\begin{thm} An entanglement witness is decomposable if and only if $\alpha_k = \alpha_{n-k}$ for $k=1,\ldots,n-1$.
\end{thm}

\begin{cor} A generic $W[\alpha_0,\ldots,\alpha_{n-1}]$ provides an indecomposable EW.
\end{cor}
Proof of the Theorem: suppose that for some $k \in \{1,\ldots,n-1\}$ one has $\alpha_k > 0 $ and $\alpha_k \neq \alpha_{n-k}$. We construct a PPT state $\rho_\epsilon$ such that $\tr (\rho_\epsilon W[\alpha_0,\ldots,\alpha_{n-1}])<0$ and hence we show that $W[\alpha_0,\ldots,\alpha_{n-1}]$ is indecomposable.
Let us consider the following  operator
\begin{equation}\label{}
    \rho_\epsilon = \Big[ \sum_{l=1}^{n-1} \Pi_l - \Pi_k - \Pi_{n-k} \Big] + \epsilon \Pi_k + \frac 1\epsilon \Pi_{n-k} + nP^+_n\ ,
\end{equation}
with $\epsilon > 0$. One easily check that both $\rho$ and $\rho^\Gamma$ are positive and hence $\rho$ represents an unnormalized PPT state.
One has
\begin{eqnarray}\label{}
\fl \tr( \rho_\epsilon W[\alpha_0,\ldots,\alpha_{n-1}] )& =& n \epsilon\alpha_k + n \frac 1\epsilon\alpha_{n-k} + n \Big(\sum_{j=0}^{n-1}\alpha_j - \alpha_k - \alpha_{n-k} \Big) - n(n-1) \nonumber \\
   &=& n \left(  \epsilon\alpha_k +  \frac 1\epsilon\alpha_{n-k} - (\alpha_k + \alpha_{n-k}) \right) \ .
\end{eqnarray}
Hence $\tr(\rho_\epsilon W[\alpha_0,\ldots,\alpha_{n-1}] ) < 0$ if $\epsilon \in (\epsilon_-,\epsilon_+)$, where
\begin{equation}\label{}
    \epsilon_\pm = \frac{ \alpha_k + \alpha_{n-k} \pm | \alpha_k - \alpha_{n-k} |}{\alpha_k} \ .
\end{equation}
It is, therefore, clear that if $\alpha_k \neq \alpha_{n-k}$, then $\epsilon_+ > \epsilon_-$ and one can always find a suitable $\epsilon$ such that $\tr(\rho_\epsilon W[\alpha_0,\ldots,\alpha_{n-1}] ) < 0$.
To prove the converse let us assume that $\alpha_k = \alpha_{n-k}$. Note that
\begin{equation}\label{}
    W[\alpha_0,\ldots,\alpha_{n-1}] = P[\alpha_0,\ldots,\alpha_{n-1}] + Q[\alpha_0,\ldots,\alpha_{n-1}]^\Gamma\ ,
\end{equation}
where
\begin{equation*}\label{}
    P[\alpha_0,\ldots,\alpha_{n-1}] = \sum_{k=1}^{n-1} \alpha_k |e_k \ot  e_{n-k} + e_{n-k} \ot e_k \>\<e_k \ot e_{n-k} +  e_{n-k} \ot e_k|\ ,
\end{equation*}
and
\begin{equation*}\label{}
    Q[\alpha_0,\ldots,\alpha_{n-1}] = \sum_{i,j=1}^{n-1} Q_{ij} E_{ij} \ot E_{ij}\ ,
\end{equation*}
where $Q_{ij}$ is a circulant matrix such that $Q_{00} = \alpha_0$ and $Q_{0k} = \alpha_k - 1$ for $k>0$. To show that $W[\alpha_0,\ldots,\alpha_{n-1}]$ is indecomposable one has to prove that
 $P[\alpha_0,\ldots,\alpha_{n-1}]$ and $Q[\alpha_0,\ldots,\alpha_{n-1}]$ are positive matrices. Positivity of $P[\alpha_0,\ldots,\alpha_{n-1}]$ is guaranteed by $\alpha_k \geq 0$. Now, the positivity of $Q[\alpha_0,\ldots,\alpha_{n-1}]$ is equivalent to positivity of a circulant matrix $Q_{ij}$. The eigenvalues of $Q_{ij}$ read
\begin{equation}\label{}
    \lambda_j = \alpha_0 + \sum_{k=1}^{n-1} (\alpha_k - 1) \omega^{-jk} \ ,
\end{equation}
for $j=0,\ldots,n-1$. One finds $\lambda_0=0$ and for $j > 0$
\begin{equation}\label{}
    \lambda_j  = \alpha_0 + 1 +  \sum_{k=1}^{n-1} \alpha_k \omega^{-jk} = c_{j0}+ 1\ ,
\end{equation}
where we used \eref{c-a}. Note, that condition $\alpha_k = \alpha_{n-k}$ guarantees that all $c_{j} \in \mathbb{R}$ and hence since $|c_{j0}|=1$ one has $c_{j0} = \pm 1$ and hence $\lambda_0=n-1$ and $\lambda_j \in \{0,2\}$ for $j>0$ which proves positivity of $Q_{ij}$. \hfill $\Box$

\begin{cor} $ W[\alpha_0,\ldots,\alpha_{n-1}]$ is decomposable if and only if $c_{k0} = \pm 1$ for $k=1,\ldots,n-1$.
\end{cor}

\begin{Ex} Taking $c_{k0} = -1$  one finds $\alpha_0=0$ and $\alpha_k = 1$ for $k>0$. This way one reproduces an entanglement witness corresponding to the reduction map.
\end{Ex}

If $\Lambda : M_n(\mathbb{C}) \rightarrow  M_m(\mathbb{C})$ is a linear map, then the dual map $\Lambda^\# : M_n(\mathbb{C}) \rightarrow  M_n(\mathbb{C})$ is defined by
\begin{equation}\label{}
    \tr [ A \Lambda^\# (B) ] : = \tr [ B \Lambda( A) ]\ ,
\end{equation}
for any $A,B \in M_n(\mathbb{C})$. If $W$ is a bipartite operators corresponding to $\Lambda$ via \eref{CJ}, then denote by $W^\#$ an operator  corresponding to $\Lambda^\#$. One finds
\begin{equation}\label{}
    W[\alpha_0,\alpha_1\ldots,\alpha_{n-1}]^\# = W[\alpha_0,\alpha_{n-1},\ldots,\alpha_{1}]\ .
\end{equation}
Interestingly, one has the following relation
\begin{equation}\label{}
    W[\alpha_0,\alpha_1\ldots,\alpha_{n-1}]^\# = \mathbb{F}\, W[\alpha_0,\alpha_{1},\ldots,\alpha_{n-1}]\, \mathbb{F}\ ,
\end{equation}
where $\mathbb{F}$ denotes a flip operator.

\begin{cor} An entanglement witness $ W[\alpha_0,\alpha_1\ldots,\alpha_{n-1}]$ is decomposable if and only if
$W[\alpha_0,\alpha_1\ldots,\alpha_{n-1}]^\# =  W[\alpha_0,\alpha_1\ldots,\alpha_{n-1}]$ or, equivalently, if
the corresponding positive map $\Lambda$ is self-dual, i.e. $\Lambda^\# = \Lambda$.
\end{cor}
Interestingly, for $n=3$ it was shown \cite{O1,O2} that if $a\leq 1$, then $W[a,b,c]$ provides a set of optimal witnesses. Optimality of $ W[\alpha_0,\ldots,\alpha_{n-1}]$ for $n > 3$ deserves further studies.

\section{Conclusions}

We analyzed a class of Bell diagonal entanglement witnesses displaying an additional $G_1 \ot G_1^*$-symmetry.
This class is characterized by a set of parameters $\{\alpha_0,\ldots,\alpha_{n-1}\}$ satisfying a family  of conditions. Interestingly, when
transformed via discrete Fourier transform it gives rise to a family of complex coefficients $c_{k0}$ satisfying remarkably simple conditions, that is, $|c_{k0}|=1$ for $k=1,\ldots,n-1$. It proves that the family of entanglement witnesses is characterized by a torus $\{\phi_1,\ldots,\phi_m\}$, where $c_{k0} = e^{i\phi_k}$ and $m = [n/2]$. Actually, if $n$ is odd there is only one torus, however, if $n$ is even there are two tori. Interestingly, the structure of these tori corresponds to properties of orthogonal groups -- torus provides a maximal abelian subgroup of $SO(n-1)$. Finally, we showed that a generic element from the class defines an indecomposable entanglement witness. Optimality of $ W[\alpha_0,\ldots,\alpha_{n-1}]$ for $n > 3$ provides an interesting open problem.

\section*{Acknowledgements}

This paper was partially supported by the National Science Center project DEC-
2011/03/B/ST2/00136. I thank Andrzej Kossakowski and Gniewko Sarbicki for
discussions and Dorota Chru\'sci\'nska for producing figures.

\end{document}